**UNIVERSIDADE FEDERAL DE SÃO CARLOS**
CENTRO DE CIÊNCIAS EXATAS E DE TECNOLOGIA
DEPARTAMENTO DE ENGENHARIA DE PRODUÇAO

PEDRO CÉSAR LOPES GERUM

# MODELAGEM DE UM PROBLEMA DE DIMENSIONAMENTO DE LOTES COM DEMANDA VARIÁVEL E DETERMINÍSTICA E EFEITOS DE *LEARNING* E *FORGETTING*

São Carlos

2013

**UNIVERSIDADE FEDERAL DE SÃO CARLOS**
CENTRO DE CIÊNCIAS EXATAS E DE TECNOLOGIA
DEPARTAMENTO DE ENGENHARIA DE PRODUÇAO

# MODELAGEM DE UM PROBLEMA DE DIMENSIONAMENTO DE LOTES COM DEMANDA VARIÁVEL E DETERMINÍSTICA E EFEITOS DE *LEARNING* E *FORGETTING*

Projeto de Pesquisa apresentado como parte dos requisitos para a realização da Monografia do Curso de Graduação em Engenharia de Produção da Universidade Federal de São Carlos (UFSCar).

**Orientando : Pedro César Lopes Gerum**
**Orientador: Prof. Dr. Roberto Fernandes Tavares Neto**

São Carlos
2013

PEDRO CÉSAR LOPES GERUM

**MODELAGEM DE UM PROBLEMA DE DIMENSIONAMENTO DE LOTES COM DEMANDA VARIÁVEL E DETERMINÍSTICA E EFEITOS DE *LEARNING* E *FORGETTING***

Monografia aprovada como parte dos requisitos para obtenção do título de bacharel em Engenharia de Produção pela Universidade Federal de São Carlos.

**Banca Examinadora**

**Orientador :**  ________________________________________
Prof. Dr. Roberto Fernandes Tavares Neto
Universidade Federal de São Carlos (UFSCar)

**Examinador :**  ________________________________________

São Carlos, ____ / ____ / ____

Dedico este trabalho à Deus, pelo apoio constante e inesgotável; aos meus pais por sempre acreditarem em mim e estarem presente; aos meus amigos pela motivação em continuar sempre e ao meu orientador por me guiar neste trabalho e dividir seu conhecimento.

# RESUMO

LOPES GERUM, P. C. **Modelagem de um problema de dimensionamento de lotes com demanda variável e determinística e efeitos de *learning* e *forgetting***

O objetivo desta pesquisa foi analisar a importância que os efeitos de *learning* e *forgetting* podem ter em um problema de dimensionamento de lotes. Tal estudo baseou-se no conceito de que a curva de aprendizagem e o ganho em escala estão presentes em muitas indústrias e são, em sua maioria, relevados no estudo de dimensionamento de lotes.

A importância de tais efeitos foi demonstrada e quantificada mostrando que há espaço para ganhos nesta área. Entretanto, por ser um problema quadrático, existe a possibilidade de que ainda haja melhoras a ser feitas nos algoritmos para que um resultado mais otimizado e robusto seja encontrado. Os resultados gerais encontrados com os algoritmos atuais, porém, mostram que a contribuição de um desconto de aprendizagem pode ser significativo, mesmo sendo em valores muito baixos.

**Palavras-chave:** Pesquisa Operacional, Dimensionamento de Lotes, Curva de Aprendizagem

# ABSTRACT

GERUM, P. C. L. **Modelling a lot-sizing problem with variable and deterministic demand and effects of learning and forgetting**

The main goal of this paper was to analyze the importance that the effects of learning and forgetting might have in a lot-sizing problem. It assumes that the learning curve and the economies of scale are present in several industries yet are, in most cases, not considered when dealing with a lot-sizing problem.

The importance of the effects was demonstrated and quantified, showing that there is still space for developments in this field. However, as the problem becomes quadratic, there is a possibility that the current algorithms are not able to solve the problem to optimality. Thus, future improvements in the algorithms may further improve the results. However, the overall results found with current algorithms show that the contribution of a discount from a learning curve can be very considerable, even if it is a minimal amount.

**Key Words:** Operational Research, Lot-Sizing, Learning Curve

# LISTA DE SÍMBOLOS

$p_i$    Tamanho do lote no período $i$

$D_i$    Demanda no período $i$

$A_i$    Custo de *setup* de uma rodada de produção no período $i$

$h_i$    Custo de estocar um produto do período $i$ ao período $i+1$

$uc_i$   Custo unitário do produto

$y_i$    Binário que define se o custo de *setup* será ou não computado no período

$C_i$    Capacidade de produção no período $i$

M       Big M, um número grande para servir de limitação superior ao modelo

$I_i$    Estoque no final do período $i$

$x_i$    Desconto unitário aplicado no custo de produção de um produto a mais no período

$t_{Ai}$ Tempos de *setup*

$b_i$    Tempo de produção de um produto no período $i$.

# LISTA DE GRÁFICOS

# LISTA DE TABELAS

# SUMÁRIO

# 1 INTRODUÇÃO

Problemas de dimensionamento de lotes são de fundamental importância para as áreas de manufatura e produção, devido ao enorme impacto direto na lucratividade que soluções ótimas trazem. Mesmo soluções heurísticas já se mostram de grande valia para as empresas, trazendo lucros ou redução de custos consideráveis. Além disso, há vasto campo nas indústrias para melhor entendimento destes problemas devido à sua recenticidade e sua pequena otimização no campo industrial. Primeiramente foram estudadas soluções empíricas e, portanto restritas. A indústria começou a demandar soluções mais abrangentes que mais áreas pudessem utilizar, importando menos o detalhamento de cada uma.

Desta forma, vários estudos foram e têm sido feitos sobre tal assunto, usando diferentes suposições para tentar se adaptar as diferentes especifidades de cada indústria. O exemplo mais clássico de tais problemas é o do Lote Econômico, ou *Economic Lot Sizing*, proposto por Ford W. Harris em 1913. Nele, assume-se uma máquina e um produto único. Suas suposições levam em conta que a demanda e os tempos de processamento são constantes e conhecidos. (LI-YAN et al, 2009; RUSSELL E TAYLOR, 2003).

De acordo com Gonzalez e González (2010), atualmente muitas empresas sofrem com um desconhecimento de um método de modelagem de *dimensionamento de lotes* e utilizam métodos que muitas vezes não se adaptam às características daquela empresa. Mesmo o modelo mais simples do lote econômico é, muitas vezes desconhecido das empresas. Mesmo aquelas que o conhecem e optam pela sua simplicidade, tem por vezes uma demanda não constante nem conhecida. Usar um modelo não apropriado acaba por trazer mais problemas do que benefícios, gerando um custo maior para a empresa.

Empresas tendem a não conhecer a demanda e tem previsões que nem sempre são precisas. Portanto, muitos estudos tem sido feitos para melhor adequar os modelos às características de cada empresa. Uma dessas características é que o tempo de processamento varia. Segundo Pinedo (2010), quanto mais uma pessoa faz um trabalho, mais rapidamente ela tende a fazê-lo. Paralelamente, uma tarefa que tem uma rotatividade muito alta sofre um efeito oposto. Esses fenômenos são conhecidos com *learning* e *forgetting* respectivamente. Ultimamente, considerações sobre o fenômeno de *learning* e *forgetting* têm sido levadas em conta em diversas pesquisas e artigos tal como o de Cheng et al (2004) e Biskup (2008).

Alguns autores mais recentes utilizaram métodos heurísticos e o método *branch and bound* para resolver tais problemas de dimensionamento de lotes. Em tal método, um fator (peso) foi dado ao tempo de processamento neste caso, como mostrado por Wu, Lee e Shiau (2006). Outros consideraram um processo envolvendo duas máquinas organizadas em *flowshop* e consideraram que os efeitos de *forgetting* são lineares. (SHIAU et al, 2007)

Apesar da extensa literatura sobre os efeitos de *learning* e *forgetting*, poucos são os estudos que aplicam tais efeitos em problemas de dimensionamento de lotes. Os poucos estudos que os aplicam, raramente consideram que os dois aconteçam simultaneamente em um processo.

Na vida real, porém, tais efeitos podem sim ser encontrados em paralelo. Em um mundo que se torna mais competitivo a cada dia, há uma forte preocupação em dar aos consumidores uma maior variedade de produtos. Consequentemente, a empresa acaba por ter bastante trocas de produtos e de times de funcionários. Dessa forma, os efeitos são maximizados dado que os trabalhadores passam por um maior número de tarefas até se tornar proficiente nelas. Com isso, há um aprendizado constante de cada pessoa que faz com que o tempo de processamento por ela realizada em determinada tarefa seja mais rápido quão maior for o tempo em que ela lá fique designinada – o efeito de *learning*. Ao mudar de posição, a tarefa sofre com o efeito de *forgetting*. Tais efeitos geram uma redução no custo que aumenta de acordo com a escala de produção. Quanto maior for o tamanho do lote, maior será esta curva de aprendizagem.

Este trabalho busca, portanto, verificar a possibilidade de aplicação de forma integrada dos efeitos estudados de *forgetting* e *learning* em outros artigos. O estudo se baseará em um caso de dimensionamento de lotes simplicado, supondo uma produção de máquina e dois produtos com demanda variável e conhecida. O objetivo do modelo é minimizar o custo total, de estoque e *setups*, sem permitir que a demanda não seja atendida.

**1.1 Problema de Pesquisa**

Com certas suposições, em um sistema de produção com produto único e máquina única, quais devem ser os pontos de refazer o pedido de cada produto no sistema e de quanto este pedido deve ser para que o custo total de produção seja minimizado, em um número determinado de períodos, levando em consideração efeitos de *learning* e *forgetting* e uma demanda variável e determinística?

**1.2 Objetivo de Pesquisa**

Criar um modelo simplificado de dimensionamento de lotes que determine os pontos de refazer o pedido de cada produto no sistema e de quanto este pedido deve ser para que o custo total de produção seja minimizado em um sistema de produção com produtos único e máquina única, em um número determinado de períodos, levando em consideração efeitos de *learning* e *forgetting* e uma demanda variável e determinística.

## 2 REVISÃO BIBLIOGRÁFICA

Todos os bens e serviços que existem ao nosso redor são produzidos em um sistema de que é organizado por um sistema de produção. Segundo Slack et al (1997), um sistema de produção é melhor otimizado ao se tomar decisões de maneira eficaz de forma que forneça:

- fácil fluxo de consumidores;
- ambiente limpo e bem projetado;
- bens suficientes para satisfazer a demanda;
- funcionários suficientes para atender os consumidores e repor os estoques;
- qualidade apropriada de serviços;
- fluxo contínuo de ideias para melhorar o desempenho de suas operações.

Este trabalho lidará primordialmente com o terceiro item: bens suficiente para atender a demanda.

De acordo com Pinedo (2010), um sistema de produção funciona através de ordens de produção. São decisões que definem quando e quanto deve ser produzido de cada produto. Essas decisões devem ser tomadas para que a demanda seja atingida e nenhum cliente seja perdido. Entretanto, comumente há várias combinações de possibilidades de ordens que atendam a demanda. O problema de dimensionamento de lotes busca encontrar a combinação que gere o menor custo para a empresa.

Tal problema pode ser lidado de três formas diferentes, no curto, médio e longo prazo. Segundo Slack et al (1997), no caso de longo prazo, a ênfase dada está mais ligada ao planejamento estratégico que à operação em si. A previsão de demanda não vai ser muito precisa e é comumente utilizada de forma agregada. O planejamento de médio prazo lida com uma demanda um pouco mais desagregada que o de longo prazo, já colocando um pouco mais de detalhes operacionais. Finalmente, o planejamento de curto prazo, a demanda já é totalmente desagregada. Nesto ponto, muitos detalhes já estão definidos e se torna, portanto complicado fazer mudanças em larga escala. Nesta monografia, será utilizado o planejamento de curto prazo, com uma previsão de demanda já definida para os próximos 6 períodos.

É importante perceber que caso a taxa de produção ou de recebimento de produtos seja maior que a consumida pela demanda, ocorrerá uma geração de estoque. O resultado pode ser visto no Gráfico 1. Percebe-se que o tempo entre o pedido e a produção total não é imediato. Desta forma, tal como acontece na vida real, a produção começa e vai sendo consumida aos poucos. Como a taxa de produção é maior que o consumo, ocorre a geração de estoque. Uma vez que todo o pedido foi produzido (ou entregue), o estoque começa a ser consumido e seu nível começa a diminuir até chegar a zero, quando acontece o re-pedido.

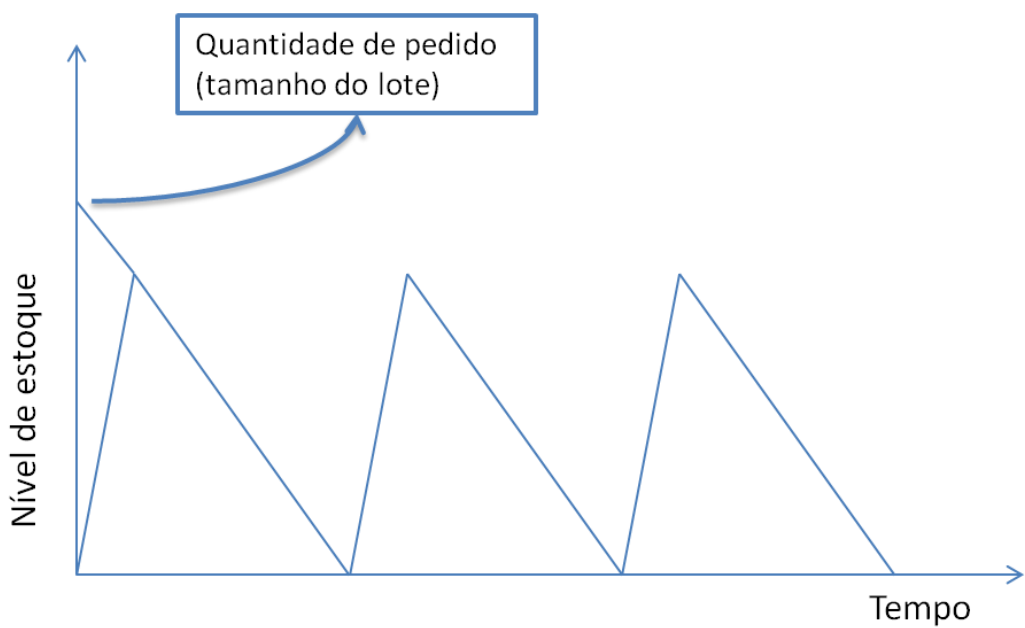

**Gráfico 1 - Perfil de estoque para reabastecimento gradual de estoque**

Fonte: Slack et al (1997), adaptado

## 2.1 Modelos para resolução de problemas de dimensionamento de lotes

Segundo Pinedo (2010), dimensionamento de lotes é um "processo de decisão que é usado regularmente em muitas indústrias manufatureiras e de serviços. Ele lida com a alocação de recursos para tarefas durante períodos de tempo dados e tem como objetivo otimizar uma ou mais variáveis." Um lote pode ser assumido como sendo a quantidade pedida de um certo produto para um fornecedor ou mesmo uma quantidade de produtos produzidos. Apesar de serem casos distintos, muitas vezes eles se interceptam na literatura devido a similaridade dos *inputs* utilizados.

O caso mais comum tem sem dúvidas o custo total como variável a ser minimizada. Neste caso do processo, a modelagem tem como premissa o fato de que os três principais custos de uma produção são:

1. Custo de *setup*: os custos fixos necessários à preparação de uma rodada de fabricação.
   - Exemplos: mão de obra diretamente aplicada na preparação das máquinas; custos dos materiais e acessórios envolvidos na preparação; outros custos indiretos: administrativos, contábeis, etc.

2. Custo variável de fabricação de cada produto: Nesse item são considerados os custos dos insumos básicos diretamente empregados no processo produtivo.
   - Exemplos: matérias primas; mão de obra diretamente aplicada na produção, tais como limpeza de máquinas, operadores de máquinas; custo de tempo de máquinas envolvidos.

3. Custo de estoque: a posse do estoque tem um custo que, para a indústria, é bastante significativo e normalmente considerado para cada produto por unidade de tempo de armazenagem.
   - Exemplos: juros de capital imobilizado; risco de obsolescência do produto; prêmios de seguro, taxas e impostos; perdas por deterioração; despesas com instalações, aluguéis, iluminação.

Os problemas de dimensionamento de lotes tem, portanto, como objetivo minimizar a somatória destes três custos, escolhendo datas e volumes para os lotes de forma que toda a demanda seja atendida.

$$Custo\ total = C_{setup} + C_{estoque} + C_{unitário}$$

Na literatura, há métodos para resolver problemas de dimensionamento de lotes especiais, com suposições consideradas para facilitar a aplicação de uma modelagem mais simplificada e fácil de ser entendida.

O método mais básico de dimensionamento de lotes assume que toda a demanda seja atendida, sem que hajam *backorders*, ou pedidos não atendidos. É simples perceber também que o custo mínimo é aquele que gera um estoque zero no final do período de estudo. Assim, em qualquer sistema de escolha de pontos de reordem e quantidades para lotes, o número de unidades a ser produzido será sempre o mesmo. Logo, para facilitar o problema, resume-se o objetivo como minimizar

$$Custo = C_{setup} + C_{estoque}$$

É possível perceber que os custos de *setup* e de estoque são portanto um *tradeoff.* Quando um aumenta, o outro diminui. Cada sistema de produção vai ter o seu ponto de tamanho de lotes e datas de colocação de ordens diferente pois dependem de características intrínsecas ao mercado em que estão atuando. Produtos que tenham alto valor agregado podem ter um custo de estoque muito baixo comparando com o custo de *setup.* Neste caso, por exemplo, uma produção de maior volume de uma só vez pode ser mais vantajosa do que uma linha de produção que recebe várias ordens em momentos diferentes.

Fernandes e Godinho (2009) ressaltam que o melhor caso para uma empresa seria buscar o custo e tempo de *setup* igual a zero, permitindo assim a utilização de um lote unitário. Entretanto, tal conceito só funciona na teoria e, praticamente, há sempre um custo de *setup* envolvido quando se há troca de produtos em uma linha ou quando a produção não é contínua e há pausas. Mesmo tendo custo muito pequeno, um tempo de *setup* muito alto também geraria um problema grande para a utilização de um lote muito pequeno. Entretanto, para efeitos de cálculo e modelagem, neste trabalho, assume-se que o tempo de *setup* não influencia o problema.

### 2.1.1 Lote Econômico de Produção (EOQ)

Um dos principais modelos de resolução do problema de dimensionamento de lotes é o do lote econômico de produção, ou *economic order quantity (EOQ).* É um dos modelos mais básicos e simples e foi um dos primeiros a ser desenvolvido para o caso. A primeira aparição do modelo data de 1913, onde Ford W. Harris foi o primeiro a propor tal solução, na revista *Factory.* Entretanto, segundo Hax e Candea (1984), o maior divulgador do modelo foi R. H. Wilson, um consultor e tem boa parte do crédito dado a ele.

O modelo do lote econômico de produção tem sete suposições:
1. A compra de matéria-prima e a produção são instantâneas;
2. Não há restrição de capacidade;
3. Demanda é determinística;
4. Demanda é constante;
5. Uma ordem de produção gera um custo de *setup* fixo, sem importar o tamanho do lote;
6. Produtos podem ser analisados individualmente;
7. *Backorders,* ou seja, não-atendimento da demanda, não são permitidos.

Neste caso, não necessariamente precisa-se saber o custo unitário do produto. Entretanto, é interessante sabê-lo para poder calcular o custo total de produção

O Gráfico 2 mostra, na prática, que o custo de setup sobe exponecialmente quanto menor for o tamanho do lote, enquanto que o custo de estoque sobe linearmente. O que se busca neste método é o ponto onde o tamanho de lote tem um custo total mínimo. Este gráfico é assim para o caso do lote econômico; em casos com outras suposições, o tamanho de lote mínimo será variável, pois a demanda também não é constante.

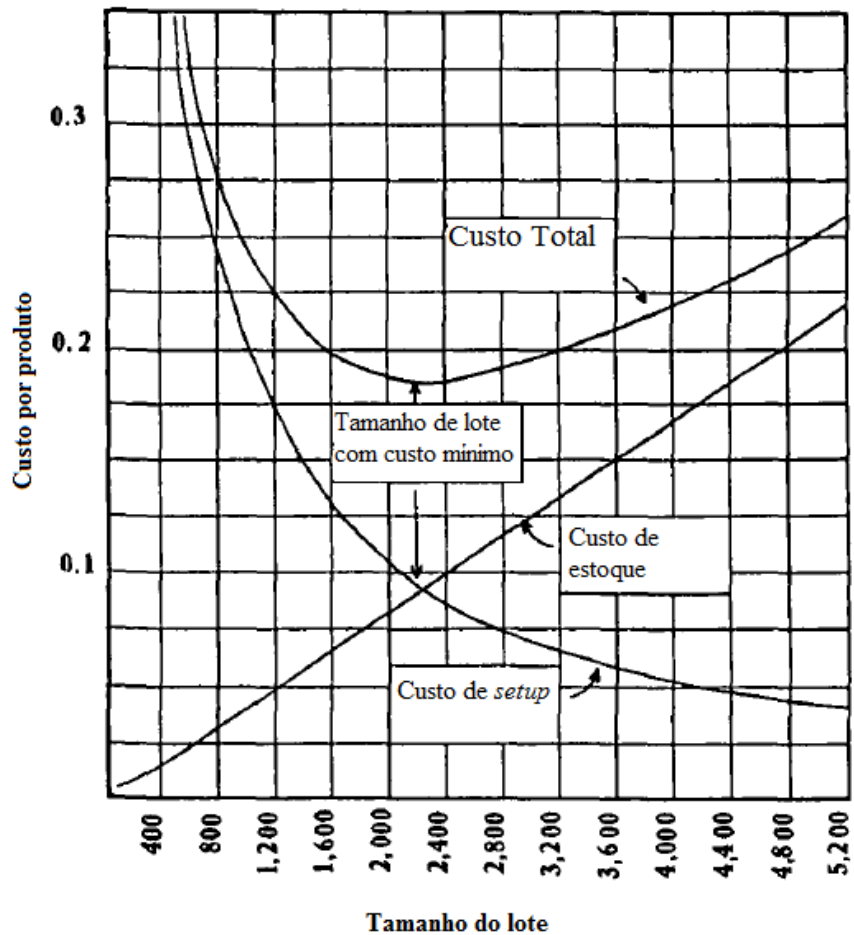

**Gráfico 2- Custos total, de estoque e de *setup***

Fonte: Harris, F. W. *Factory, the magazine of management*, volume 10, número 2, p. 136, 1913

O Gráfico 3 (SLACK et al, 1997, adaptado) mostra o comportamento do estoque utilizando o modelo EOQ. O gráfico também mostra como funcionaria se a suposição 1 não existisse, mostrando que o ponto de reordem depende do *lead time* ou tempo de fornecimento. Segundo Lambert et al (1998), é o período entre o início de uma atividade, produtiva ou não, e o seu término. Neste caso, *lead time* significa o tempo entre o momento em que o pedido é feito e o momento em que os produtos estão prontos para serem vendidos. Como neste modelo, consideramos o *lead time* como sendo igual a zero, o ponto de reordem é aquele aonde o gráfico corta o eixo do tempo.

É importante perceber que a linha de estoque desce linearmente, caracterizando uma demanda constante, o que caracteriza a suposição 4. O cálculo do custo de estoque leva em conta a quantidade média de estoque. Essa quantidade pode ser encontrada facilmente e é mostrada no gráfico e é igual a metade do tamanho do lote.

Olhando o Gráfico 3, percebemos que sua principal diferença em relação ao Gráfico 1 é que a reposição do estoque é realizada em um único momento por pedido. Assim, ocorrem picos nos momentos onde o pedido é entregue. Tal suposição 1 facilita muito os cálculos e isso é um dos motivos pelos quais esse método é muito difundido e conhecido.

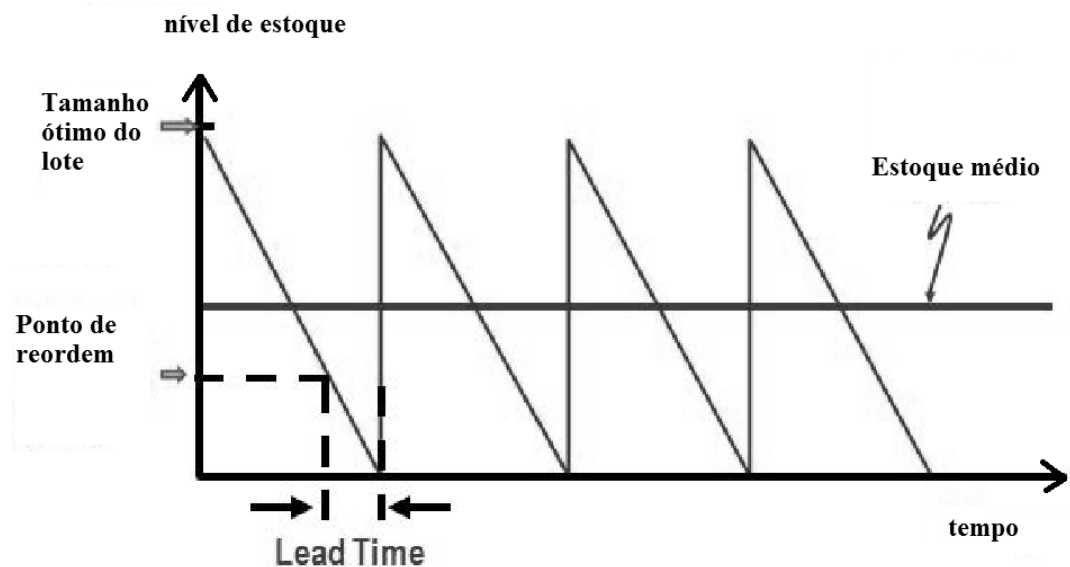

**Gráfico 3 - Relação do nível de estoque com o tempo, utilizando o modelo do lote econômico**
Fonte: http://www.investigaciondeoperaciones.net/eoq.html, visitado em 04/07/2013, às 21:03

Neste modelo, os *inputs* utilizados são:

1. Demanda, que é constante e determinada;
2. Custo de estoque de um produto em um período;
3. Custo de *setup*, ou seja, de fazer o pedido ou começar a rodar a produção;

Abaixo, seguindo a demonstração do modelo, Y significa o custo total para se produzir/comprar uma quantidade Q de produtos; D é a demanda constante e determinada que ocorre todo período; c é o custo unitário de produção e h é o custo de estocar um produto por um período.

Para se encontrar o número de ordens necessárias para atender a demanda em um período, divide-se a demanda total pelo tamanho do lote. É fácil perceber que o tempo que leva para se esgotar o estoque é o inverso deste número de ordens. Logo temos

$$Custo\ total(p) = \frac{AD}{p} + cD + \frac{ph}{2}$$

que representa a soma dos três principais custos de produção, que são, respectivamente, custo de *setup*, custo variável da fabricação de cada produto e custo de estoque. O modelo tem o objetivo de buscar o *p* que minimize *Custo total(p)*. Para tanto, deriva-se a fórmula e a iguala a zero. Com isso, temos que

$$p_{ótimo} = \sqrt{\frac{2AD}{h}}$$

Repare que o custo variável de fabricação não interfere no cálculo para o tamanho de lote que gera o custo mínimo, pois ele é o mesmo, independente de qual for o tamanho deste lote. Finalmente, substituindo a quantidade do lote encontrado, obtemos

$$Custo\ total_{mínimo} = \sqrt{2ADh} + cD$$

**2.1.2 Dimensionamento de lotes dinâmicos**

Quando a demanda não é constante, foge-se das suposições de um modelo como o do lote econômico. Para se resolver um problema com demanda determinística, mas variável, é necessário dividir a demanda em períodos de tempo *i* que pode significar dias, semanas ou até meses, dependendo do tamanho do volume do sistema.

Com isso, não há um tamanho de lote definido que minimiza o custo, mas sim um conjunto de tamanho de lotes e de pontos de refazer o pedido que geram um custo total final mínimo. As outras suposições continuam as mesmas do modelo EOQ.

Devido a dificuldade de se determinar os lotes que produzem um custo mínimo, vários procedimentos de fácil entendimento e aplicação foram propostos na literatura, tal como a regra do lote por lote e a regra da quantidade de pedido fixa. Além destes, o algoritmo de Wagner Whitin foi desenvolvido para buscar a solução ótima de problemas de dimensionamento de lotes deste tipo.

**2.1.2.1 Regra lote por lote**

É o procedimento mais simples que consiste em produzir exatamente o que se pede no período *i*. Portanto, o custo de estoque é zero e o custo total é dado pelo custo de *setup* multiplicado pelo número de períodos. Se por exemplo $i = 1, 2, 3, ..., j$, chegamos a conclusão de que:

$$Custo = Aj$$

$$Custo\ total = Aj + uc * \sum_i D_i$$

Logo, os únicos *inputs* necessário para tal modelo é
1. O custo de *setup* de cada período;
2. O custo unitário de produção/compra;
3. A demanda de cada período.

**2.1.2.2 Regra da quantidade de pedido fixa**

Esta regra tem por característica pedir sempre a mesma quantidade, tal como era feito no modelo do EOQ, mas com a diferença que, neste caso, os pontos de refazer o pedido são variados. Esta quantidade pode ser definida dependendo da capacidade de produção da empresa e de sua capacidade de estoque. Assim, há necessidade de um *input* extra, que seria o custo de estocar um produto em um período.

**2.1.2.3 Algoritmo de Wagner Whitin**

O algoritmo de Wagner Whitin foi criado por Harvey M. Wagner e Thomson M. Whitin em 1958 e encontra soluções ótimas para problemas do tipo de lotes dinâmicos. Ele é muito utilizado como *benchmark* de outros modelos pois encontra a solução ótima, permitindo que se verifique quão boa a outra solução é. No entanto, por ser um modelo que depende de recursividade, ele acaba tendo alta complexidade matemática iterativa quando o número de variáeis não é razoavelmente pequeno. Logo, ele não é o principal modelo utilizado, dando lugar a outros modelos mais simples para o dia a dia das empresas.

Segundo Gonçalves (2000), há duas propriedades que a solução deve satisfazer:

1. Um pedido só pode ser feito quando o nível de estoque atinge zero. Isso quer dizer que ou o estoque de *i-1* é igual a zero, ou o custo de *setup* do período *i* é igual a zero de forma que

$$A_i * I_{i-1} = 0$$

2. Existe um limite superior para o número de períodos para os quais uma encomenda durará. Isso quer dizer que

$$p_i \leq \sum_{i=1}^{j} D_i$$

O modelo pode ser divido em várias etapas, cada uma delas representando um dos períodos que englobam o total. Logo, com $i = 1, 2, 3, ..., j$, precisamos de *j* iterações para encontrar a solução ótima. Como exemplo, suponhamos que temos 3 períodos ($j = 3$).

Passo 1: obrigatoriamente temos que produzir no período 1 para atender a demanda. Logo, negligenciando o custo variável de produção de cada produto temos que:

$$Custo_{ótimo,1} = A_1$$

Passo 2: ao se aumentar o horizonte de produção para dois períodos, tem-se que verificar 2 opções. Pode-se produzir tudo no período 1 e carregar a demanda do segundo período como estoque ou produzir a demanda do período 2 neste segundo período.

$$Custo_{ótimo,2} = \min \{Custo_{ótimo,1} + A_2, A_1 + h_1 D_2\}$$

Passo 3: continuando a aumentar o número de períodos, temos três possibilidades. Pode-se produzir tudo no período 1 e carregar estoque durante os períodos subsequentes, pode-se produzir a demanda do terceiro período somente no terceiro período ou pode-se produzir as demandas do período 2 e 3 no período 2.

$$Custo_{ótimo,3} = \min\{A_1 + h_1 D_2 + (h_1 + h_2)D_3, Custo_{ótimo,1} + A_2 + h_2 D_3, Custo_{ótimo,2} + A_3\}$$

Como pode ser visto, para cada período a mais que é colocado no modelo, uma possibilidade a mais é colocada na iteração final, deixando o modelo cada vez mais complexo. Porém, uma propriedade pode ajudar a diminuir o número de iterações necessários. Se o último período em que houve produção em um passo for *i* para que o custo naquele passo fosse mínimo, só é necessário verificar os períodos subsequentes no passo seguinte.

**2.4 Os efeitos de *learning* e *forgetting***

Hill (2011) define *learning curve* como um modelo matemático que relaciona uma variável de desempenho ao número de unidades produzidas, de forma que elas sejam inversamente proporcionais. Logo, a variável de desempenho diminui enquanto a produção cumulativa aumenta.

O termo *forgetting* caracterizaria um esquecimento do aprendizado uma vez que o período se acaba. Neste trabalho, tais conceitos serão incuídos no custo variável de fabricação de cada produto, tal como mostrado no Gráfico 4. Fazendo uma suposição com a realidade, o custo de produzir cada produto diminui quanto maior for o lote produzido. Assim, esse terceiro tipo de custo passa a depender do tamanho do lote e deve, portanto, entrar no modelo proposto para ser otimizado.

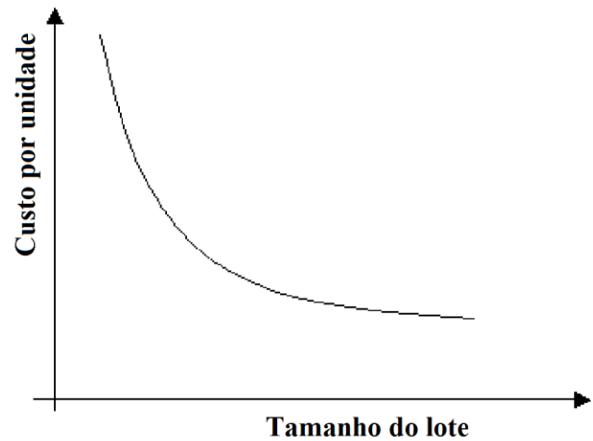

**Gráfico 4 - Efeitos de *learning* na produção de um lote em um período**
Fonte: Elaborado pelo autor

Devido às restrições computacionais e às dificuldades de modelagem, neste trabalho será utilizada a suposição de que tal efeitos seguem um padrão linear e não quadrático. Assim, assumir-se-á que o gráfico siga o padrão do gráfico 5.

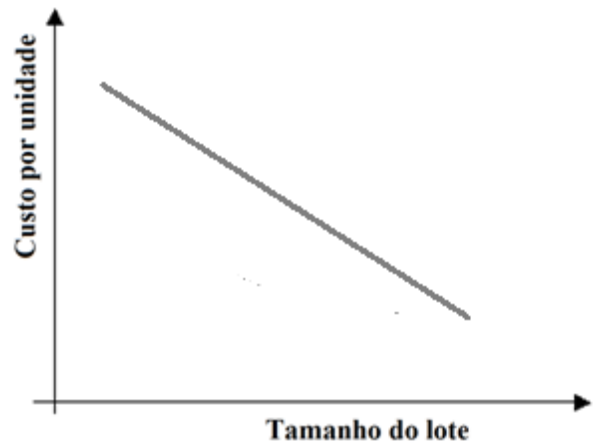

**Gráfico 5 - Efeitos de *learning* na produção de um lote em um período assumido neste trabalho**
Fonte: Elaborado pelo autor

## 2.5 A pesquisa operacional

De acordo com a Sociedade Brasileira de Pesquisa Operacional, "a Pesquisa Operacional é uma ciência aplicada, voltada para a resolução de problemas reais. Tendo como foco a tomada de decisões, aplica conceitos e métodos de outras áreas científicas para concepção, planejamento ou operação de sistemas para atingir seus objetivos.Através de desenvolvimentos de base quantitativa, a Pesquisa Operacional visa também introduzir elementos de objetividade e racionalidade nos processos de tomada dedecisão, sem descuidar, no entanto dos elementos subjetivos e de enquadramento organizacional que caracterizam os problemas".

A Pesquisa Operacional foi inicialmente concebida durante a Segunda Guerra Mundial, quando um grupo de cientistas foi convocado na para resolver problemas associados à defesa da Inglaterra. Eles queriam que os recursos militares escassos não fossem utilizados de forma impensada e buscavam maneiras de otimizar sua utilização.

De acordo com Shariful (2002), seguindo o exemplo inglês, os EUA começaram o estudo de Pesquisa Operacional inicialmente dentro da academia e deve-se muito à ampla propagação de tal conhecimento, especialmente à equipe liderada por George B. Dantzig que originou aquele que viria a ser conhecido como método Simplex. Após a guerra, a utilização de Pesquisa Operacional começou a se expandir para áreas mais diversas aumentando a abrangência e complexidade dos problemas encontrados.

### 2.5.1 Utilização de pesquisa operacional em problemas de dimensionamento de lotes

A complexidade inerente de problemas de dimensionamento de lotes como visto acima, muitas vezes, requer ajuda computacional para ser resolvido. A pesquisa operacional vem como ajuda neste caso, buscando modelar os problemas matematicamente.

De acordo com Arenales et al (2007, p. 4):

> A partir da observação de fenômenos, processos ou sistemas, [...] buscam-se leis que os regem. Essas leis, se passíveis de serem descritas por relações matemáticas, dão origem aos *modelos matemáticos*. O termo *modelo* [...] é usado para como objeto abstrato, que procura imitar as principais características de um objeto real para fins de representar o objeto real.

Para facilitar a geração de soluções, é comum a simplificação do problema para possibilitar uma modelagem mais rápida e que permita computadores e *softwares* mais triviais de resolvê-lo. Normalmente, tenta-se deixar o problema de forma linear, para que métodos robustos de programação linear possam solucioná-lo de forma rápida e eficaz.

**2.5.1.1 Programação Linear**

Um modelo de Programação Linear é um modelo matemático de otimização no qual todas as funções são lineares (tanto a função objetivo, quanto as restrições). De acordo com Goldbarg e Luna (2000), um modelo de Programação Linear deve possuir as seguintes características:

• Proporcionalidade: a quantidade de recurso consumido por uma dada atividade
deve ser proporcional ao nível dessa atividade na solução final do problema.
Além disso, o custo de cada atividade é proporcional ao nível de operação da
atividade;
• Não Negatividade: deve ser sempre possível desenvolver dada atividade em
qualquer nível não negativo e qualquer proporção de um dado recurso deve
sempre poder ser utilizado;
• Aditividade: o custo total é a soma das parcelas associadas a cada atividade;
• Separabilidade: pode-se identificar de forma separada o custo (ou consumo de
recursos) específico das operações de cada atividade.

Arenales et al (2007) propõem o seguinte modelo para resolver um problema simples de dimensionamento de lotes:

$$\min \sum_{i=1}^{n}(A_i y_i + h_i I_i)$$

| | | |
|---|---|---|
| $I_i = I_{i-1} + p_i - D_i,$ | $i = 1, ..., n$ | (1) |
| $\sum_{i=1}^{n}(t_{Ai} y_i + b p_i) \leq C_i,$ | | (2) |
| $p_i \leq M_i y_i,$ | $i = 1, ..., n$ | (3) |
| $M_i = \min\{\frac{C_i - t_{Ai}}{b_i}, \sum_{i=1}^{n}(D_i)\},$ | $i = 1, ..., n$ | (4) |
| $p \in R_+^n,\ I \in R_+^n,\ y \in B_+^n$ | | (5) |

onde *I* representa o estoque total no período *i*; *y* assume valores binários, representando a inclusão ou não do custo de *setup* no período *i*; *M* é um *big M* que ajuda a delimitar o valor de *p*, colocando um limitante superior nas iterações a serem feitas pelo computador e *C* representa a capacidade, em tempo de produção. $t_{Ai}$ e *b* são respectivamente os tempos de *setup* e de produção de um produto no período *i*.

A restrição 1 é a que calcula o novo estoque total período a período. É simples entender que o estoque do período atual é igual ao estoque do período anterior mais o que foi produzido menos o que foi consumido. A restrição 2 lida com a capacidade geral de produção. A terceira restrição é a que define o valor de y no período *i*. Se a produção *p* for zero, então *y* necessariamente será zero. Caso contrário, *y* assume o valor de 1. Finalmente a restrição 4 funciona como um limitante superior para permitir que menos iterções sejam feitas, gastando menos tempo e memória computacional.

Tal modelo é interessante pois permite calcular o valor ótimo ou próximo do ótimo de lotes mesmo quando os custos variam com os períodos, permitindo assim uma gama maior de utilizações.

**2.5.1.2 Programação quadrática**

Para o problema estudado neste trabalho, um exemplo de simplificação foi mostrado no gráfico 5 onde assumiu-se que os efeitos de *learning* e *forgetting* seguem um padrão linear. Mesmo com tal simplificação, torna-se necessário a utilização de um método que resolva problemas de otimização quadráticos. Isto ocorre pois o custo de produção por período será a multiplicação do número de produtos produzidos pelo seu custo unitário, que também é função do número de partes produzidas. Assim temos um problema de programação quadrática.

Um problema de programação quadrática é caracterizado por uma função objetiva quadrática e restrições lineares. Na maioria dos casos, pode ser resolvido eficientemente com ajuda computacional pois, segundo Mauri e Lorena (2009), algoritmos robustos já foram desenvolvidos para lidar com tal tipo de problemas em escala de até 200 variáveis. Métodos heurísticos tem dado bons resultados para até 2500 variáveis.

**2.6 Softwares computacionais**

Com o aumento da complexidade dos problemas, o computador se tornou uma ferramenta fundamental, aumentando a nossa capacidade de fazer cálculos e iterações. Desta forma, problemas antes insolúveis passaram a ser resolvidos. Dois dos programas de modelagem matemática visando otimização que são bastante difundidos são o *solver,* que vem junto com o *Microsoft Office Excel*, *Lindo* e o *GAMS, General Algebraic Modeling System*.

**3  METODOLOGIA DE PESQUISA**

Para se determinar a abordagem de pesquisa do projeto proposto, é importante que se entenda que este trabalho parte de problema a ser resolvido através de um modelo. Tal modelo é composto de variáveis mensuráveis que são alteradas gerando um efeito que é mensurado através da *função objetivo*.

Esta pesquisa tem como característica a replicabilidade, podendo ser usada em outras ocasiões partindo de suposições semelhantes e de dados parecidos. Neste caso, os resultados devem ser equivalentes.

Assim, observando Martins (2010), define-se esta pesquisa como quantitativa. Segundo Ferreira (apud Tavares, 2010), tal abordagem apresenta vantagens quando:
- O problema é difícil e é necessário se fazer uma abordagem quantitativa para se chegar a uma conclusão;
- O problema é facilmente automatizado, minimizando o tempo gasto para resolução.

Morabito e Pureza (2010) sugerem que a pesquisa operacional é mais efetiva quando os modelos são próximos aos processos reais e, posteriormente, testados na prática. A figura 2 exemplifica o processo que a pesquisa deve seguir. Morabito e Pureza (2010) sistematizam em 5 etapas aquilo que a figura 2 mostra: a metodologia de pesquisa em problemas quantitativos com modelagem.

As cinco etapas são:
1. definição do problema;
2. construção do modelo;
3. solução do modelo;
4. validação do modelo;
5. implementação do modelo.

## 3.1 Definição do problema

Morabito e Pureza (2010) ressaltam que é nesta fase que são definidos o escopo e os objetivos do modelo. O modelo deve ser o mais próximo possível da realidade, pois, se o modelo conceitual fugir daquilo que que é real, dificilmente resultará em uma resposta útil para o problema. O entendimento e definição do problema são, assim, partes fundamentais da pesquisa.

Nesta fase também são definidas as suposições que facilitariam a modelagem do problema matematicamente. Elas funcionam como um *tradeoff,* pois quanto mais elas simplificam o modelo, deixando-o mais fácil de ser resolvido, mais longe da realidade o modelo fica.

Este trabalho lida com uma expansão do problema de dimensionamento de lotes. A busca por uma aproximação maior à realidade mostra que os efeitos de *learning* e *forgetting* podem e devem ser incluídos neste problema. Para facilitar a implementação e resolução do problema, assumiu-se que os efeitos tem caráter linear. O problema foi definido como sendo de produto único e máquina única, com demanda conhecida e variável.

**3.2 Construção do modelo**

Nesta fase tem-se a coleta de dados e informações que serão utilizados no modelo. O modelo é então desenvolvido partindo-se de relações matemáticas e lógicas de simulação. Geralmente, utiliza-se modelos já criados e divulgados na teoria como base para um modelo mais avançado e diferente.

Durante a construção do modelo, podem ocorrer mudanças necessárias devido à complexidade matemática de certa variável ou a dificuldade de se traduzir para lógica matemática certo conceito da realidade.

Tomando como base um modelo de dimensionamento de lotes desenvolvido por Arenales et al (2007), foi adicionado a variável que leva em consideração os efeitos de *learning* e forgetting. O modelo assim se torna mais abrangente. Inicialmente, supôs-se um produto único com máquina única.

No modelo utilizado neste trabalho, assumiu-se um espaço de 6 períodos e para cada período foram escolhidos valores para demanda, custo de *setup* e custo de estoque unitário. Tomando como base a vida real, cada período representaria um mês, indo de Dezembro até Maio. O produto em questão foi pensado de forma a ter uma demanda maior com o calor, férias escolares e datas comemorativas. Logo, os dois primeiros meses possuem uma demanda consideravelmente maior que os outros. O mês de maio também causa um aumento na demanda devido ao dia das mães. A demanda está melhor explicitada na tabela 1:

**Tabela 1 - Demanda durante os 6 períodos**

| Período | 1 | 2 | 3 | 4 | 5 | 6 |
|---|---|---|---|---|---|---|
| **Demanda** | 1500 | 1500 | 400 | 200 | 400 | 1000 |

Para um entendimento do impacto que o efeito de *learning* e *forgetting* tem neste problema, o modelo foi rodado 18 vezes, com três variações no custo de estocagem unitário, duas variações no custo de *setup* e três variações na curva do custo de produção unitário.

Para o custo de estocagem, assumiu-se que ele seria constante por todos os seis períodos, sendo de $1,00, $3,00, ou $5,00. Para o custo de *setup,* assumiu-se, em um caso, que ele seria constante, valendo $2000,00 por período; no outro caso, ele é variável, conforme a tabela 2.

Finalmente o custo unitário segue a seguinte fórmula:

$Custo\ Unitário = 100 - x * (partes\ produzidas)$, onde x vale 0,01, 0,001 ou 0,0001. Sabe-se que x é menor que 0,2 (pois senão todas as 5000 partes seriam produzidas de uma vez, com custo zero). Quando x for muito pequeno, a solução para o problema seria a mesma que aquela em que os efeitos de *learning* e *forgetting* não são considerados, pois o desconto poderia ser considerado como desprezível.

Para efeitos de limitação do número de iterações do modelo, foi colocado uma limitação de capacidade de 5000 partes por período. Este número foi escolhido para não permitir que descontos gerem um custo unitário negativo e por permitirem que toda a produção de todos os períodos seja feita de uma única fez, caso este seja o melhor caso.

**Tabela 2 - Custos assumidos para o modelo**

|         | Custos de estoque ($/un) | | | Custos de *setup*($) | | Desc. no custo un. ($/parte) | | |
|---------|------|------|------|------|------|--------|--------|--------|
| Período | h1   | h2   | h3   | A1   | A2   | x1     | x2     | x3     |
| 1       | 1    | 3    | 5    | 2000 | 1500 | 0.01   | 0.001  | 0.0001 |
| 2       | 1    | 3    | 5    | 2000 | 1000 | 0.01   | 0.001  | 0.0001 |
| 3       | 1    | 3    | 5    | 2000 | 2000 | 0.01   | 0.001  | 0.0001 |
| 4       | 1    | 3    | 5    | 2000 | 3500 | 0.01   | 0.001  | 0.0001 |
| 5       | 1    | 3    | 5    | 2000 | 1500 | 0.01   | 0.001  | 0.0001 |
| 6       | 1    | 3    | 5    | 2000 | 2500 | 0.01   | 0.001  | 0.0001 |

## 3.3 Solução do modelo

Depois de pronto, o modelo é colocado em forma de algoritmo para que certos *softwares* possam resolvê-lo. Neste trabalho, serão utilizados dois *softwares*. O *solver,* que vem junto com o *Microsoft Office Excel* e o *GAMS*, muito difundido entre os estudiosos de pesquisa operacional.

Normalmente, nesta fase, faz-se testes para verificar possíveis erros que existam no código e se as soluções estão adequadas àquilo que se esperava delas. Muitas vezes se faz necessário adaptações do algoritmo para permitir que os *softwares* dêem conta dos cálculo. Para isso, são feitas análises de sensibilidade e de cenários, que verificam a consistência e a robustez das soluções.

Esta é a fase mais bem definida do processo de implementação pois envolve basicamente modelos matemáticos precisos. Neste trabalho, foi utilizado o GAMS.

O modelo com os efeitos de *learning* e *forgetting* está descrito abaixo. É importante relembrar que, para resolver o problema, foi necessário a utilização de programação mista quadrática, pois o custo de produção é resultado da multiplicação de partes produzidas pelo custo unitário, que também é função do número de partes produzidas.

### 3.3.2 Modelo com o efeito de *learning* e *forgetting*

Ao modelo de programação linear definido anteriormente, uma restrição foi adicionada. A equação número 7 mostra onde o efeito de *learning* e *forgetting* foi implementado, além de um termo extra na função objetivo. Neste novo modelo, *x* representa o desconto unitário que a produção de uma parte extra geraria no custo de produção.

$\min \sum_{i=1}^{n}(A_i y_i + h_i I_i + uc_i p_i)$

| | | |
|---|---|---|
| $I_i = I_{i-1} + p_i - D_i,$ | $i = 1, \ldots, n$ | (6) |
| $\sum_{i=1}^{n}(t_{Ai} y_i + b p_i) \leq C_i,$ | | (7) |
| $uc_i = 100 - x * p_i$ | $i = 1, \ldots, n$ | (8) |
| $p_i \leq M_i y_i,$ | $i = 1, \ldots, n$ | (9) |
| $M_i = \min\{\frac{C_i - t_{Ai}}{b_i}, \sum_{i=1}^{n}(D_i)\},$ | $i = 1, \ldots, n$ | (10) |
| $uc \in R_+^n, \; I \in R_+^n, x \in R_+^n, y \in B_+^n$ | | (11) |

Tal problema é quadrático pois o termo *uc$_i$* x *p$_i$* pode ser transformado em
$(100 - x * p_i) * p_i = (100 p_i - x p_i^2)$.

# 4 RESULTADOS

Com as soluções obtidas, verifica-se se elas se adequam ao problema e resolvem aquilo que era buscado no objetivo inicial. Um experimento em um sistema real mostra se as soluções obtidas são boas, mesmo com as suposições feitas para facilitar a lógica matemática.

É importante notar que as soluções dependem diretamente da precisão do modelo. Morabito e Pureza (2010) dizem que "um modelo mais preciso, mesmo que resolvido de forma aproximada, pode ser bem mais útil que um modelo menos preciso resolvido de forma exata."

No caso do modelo feito neste trabalho, o resultado está na tabela 3 abaixo. Nela percebemos que a inclusão do efeito gerou resultado diferente daquele proposto para um modelo mais simples. Isto demonstra que o efeito de *learning* e *forgetting* influencia sim na produção, assumindo que há ganho de pelo menos 1/10 de centavo por parte a mais produzida.

Sabe-se entretanto que um modelo quadrático não é tão preciso quanto um modelo linear. Assim, deve-se dizer que os resultados abaixo não necessariamente correspondem àquilo que seria a solução ótima do problema, representando somente um mínimo local. Mesmo assim, como foi dito acima, é melhor uma solução aproximada de um modelo mais preciso que uma solução exata de um problema menos preciso.

**Tabela 3 - Resultados encontrados**

| Rodada | Tipo de custo de estoque | Tipo de custo de *setup* | Tipo de desc. Unit. | Produção | | | | | | Custo Total | |
|---|---|---|---|---|---|---|---|---|---|---|---|
| | | | | 1 | 2 | 3 | 4 | 5 | 6 | | |
| 1 | h1 | A1 | x1 | 3600 | | | | 1400 | | $ | 358,700 |
| 2 | h1 | A1 | x2 | 3600 | | | | 1400 | | $ | 492,980 |
| 3 | h1 | A1 | x3 | 3000 | | 400 | 200 | 1400 | | $ | 505,384 |
| 4 | h1 | A2 | x1 | 3600 | | | | 1400 | | $ | 357,700 |
| 5 | h1 | A2 | x2 | 3600 | | | | 1400 | | $ | 491,980 |
| 6 | h1 | A2 | x3 | 1500 | 1500 | 400 | 200 | 1400 | | $ | 504,334 |
| 7 | h2 | A1 | x1 | 3600 | | | | 1400 | | $ | 366,500 |
| 8 | h2 | A1 | x2 | 3000 | | 400 | 200 | 1400 | | $ | 500,340 |
| 9 | h2 | A1 | x3 | 1500 | 1500 | 400 | 200 | 1400 | | $ | 508,334 |
| 10 | h2 | A2 | x1 | 3600 | | | | 1400 | | $ | 365,500 |
| 11 | h2 | A2 | x2 | 3000 | | 400 | 200 | 1400 | | $ | 499,340 |
| 12 | h2 | A2 | x3 | 1500 | 1500 | 400 | 200 | 1400 | | $ | 506,334 |
| 13 | h3 | A1 | x1 | 3600 | | | | 1400 | | $ | 374,300 |
| 14 | h3 | A1 | x2 | 1500 | 1500 | 400 | 200 | 1400 | | $ | 504,340 |
| 15 | h3 | A1 | x3 | 1500 | 1500 | 400 | 200 | 1400 | | $ | 510,334 |
| 16 | h3 | A2 | x1 | 3600 | | | | 1400 | | $ | 373,300 |
| 17 | h3 | A2 | x2 | 1500 | 1500 | 400 | 200 | 1400 | | $ | 502,340 |
| 18 | h3 | A2 | x3 | 1500 | 1500 | 400 | 200 | 1400 | | $ | 508,334 |

Observando os resultados na Tabela 3, vê-se que o desconto gerado pelo efeito de *learning* e *forgetting* tem um impacto que deve ser considerado. Comparando casos em que o custo de estoque e custos de *setup* se mantiveram constantes, vemos que há variações no dimensionamento otimizado.

Dependendo do nível de desconto, a variação no custo total pode ser brutal. Isto ocorre principalmente porque a principal parte no custo é o custo de produção. Mesmo com um desconto pequeno, na faixa de 1/10 de centavo por produto produzido (o que equivale a 0,01% do custo original do produto), a média de redução geral de custos foi de $10.000. Quanto esse desconto sobe para 0,1% do preço original por produto, a redução fica em torno de $150.000.

No caso da variação de custos de *setup,* ambos os tipos A1 e A2 somam o mesmo valor final. No primeiro caso, o custo é igual para todos os períodos, mas para o caso A2, o custo é variável. Pelos resultados, é possível perceber que, neste exemplo, a variação permitiu uma melhora estratégica no dimensionamento de lotes. Essa redução de custos, entretanto, foi bem menor que aquela relacionada ao desconto de produção e ficou em torno de $1,000.

Finalmente olhando para o estoque, vemos que o aumento feito em cada diferente rodada do modelo foi bem alto. Mesmo assim, o caso em que o custo de estoque quintuplicou pode ser equiparado com um aumento no desconto, passando de $0,001 para $0,01. Na tabela, isso é mostrado nas rodadas 3 e 14, respectivamente.

# 5 CONSIDERAÇÕES FINAIS

Este trabalho teve como objetivo analisar possíveis impactos que os efeitos de *learning* e *forgetting* teriam em um problema de dimensionamento de lotes com máquina simples e produto único; a demanda e os custos foram considerados variáveis. Um modelo criado necessitou de programação quadrática para ser resolvido de forma que um apoio computacional se deu necessário. Com isso, GAMS foi escolhido.

Os resultados obtidos mostraram que os efeitos de *learning* e *forgetting* são sim importantes e que devem ser considerados se uma economia de escala for algo que esteja de acordo com a realidade. Isto ocorre devido à grande fração do custo total que é dependente do custo de produção, diretamente relacionado ao desconto de escala. Um desconto de 0,1% por produto do custo inicial de produção gerou reduções de custo de 30% no problema estudado.

Tal desconto pode provir de várias formas, seja de fornecedores ou até mesmo de um ganho de velocidade e redução de custos devido à curva de aprendizagem.

Para futuros trabalhos, uma oportunidade levantada foi a criação de um modelo que leve em conta a curvatura da curva de aprendizagem, que é mais comumente logarítmica. Uma expansão para problemas multi-máquinas e multi-produtos também é uma boa continuação ao trabalho aqui iniciado.

# REFERÊNCIAS

## APÊNDICE A - EXEMPLO DE MODELO UTILIZADO NO GAMS

sets

    i period / 1, 2, 3, 4, 5, 6 /

parameters

    d(i) demand in each period

    /    1    1500

         2    1500

         3    400

         4    200

         5    400

         6    1000  /

    C(i) capacity in each period per day

    /    1    5000

         2    5000

         3    5000

         4    5000

         5    5000

         6    5000 /

    A(i) setup in each period of the day

    /    1    4000

         2    4000

         3    4000

         4    4000

         5    4000

         6    4000/

    h(i) setup in each period of the day

```
        /   1    1
            2    1
            3    1
            4    1
            5    1
            6    1/;

scalar m Numero grande /100000000/;

variables
    p(i)   producao no periodo
    Inv(i) estoque final no periodo
    uc(i)  custo producao no periodo
    y(i)   binario
    z      custo total

Positive variable p;
Positive variable Inv;
Positive variable uc;
Binary variable y;

Equations
I1          estoque periodo 1
I2          estoque periodo 2
I3          estoque periodo 3
I4          estoque periodo 4
I5          estoque periodo 5
I6          estoque periodo 6
Cap1        capacidade periodo 1
```

| | | |
|---|---|---|
| Cap2 | capacidade periodo 2 | |
| Cap3 | capacidade periodo 3 | |
| Cap4 | capacidade periodo 4 | |
| Cap5 | capacidade periodo 5 | |
| Cap6 | capacidade periodo 6 | |
| bin1 | binario producao | |
| bin2 | binario producao | |
| bin3 | binario producao | |
| bin4 | binario producao | |
| bin5 | binario producao | |
| bin6 | binario producao | |
| costpart1 | | |
| costpart2 | | |
| costpart3 | | |
| costpart4 | | |
| costpart5 | | |
| costpart6 | | |
| cost | custo total ; | |
| I1.. | Inv('1') | =e= p('1')-d('1'); |
| I2.. | Inv('2') | =e= Inv('1')+p('2')-d('2'); |
| I3.. | Inv('3') | =e= Inv('2')+p('3')-d('3'); |
| I4.. | Inv('4') | =e= Inv('3')+p('4')-d('4'); |
| I5.. | Inv('5') | =e= Inv('4')+p('5')-d('5'); |
| I6.. | Inv('6') | =e= Inv('5')+p('6')-d('6'); |
| Cap1.. | C('1') | =g= p('1'); |
| Cap2.. | C('2') | =g= p('2'); |
| Cap3.. | C('3') | =g= p('3'); |
| Cap4.. | C('4') | =g= p('4'); |
| Cap5.. | C('5') | =g= p('5'); |
| Cap6.. | C('6') | =g= p('6'); |

```
bin1..          p('1')       =l= y('1')*m ;
bin2..          p('2')       =l= y('2')*m ;
bin3..          p('3')       =l= y('3')*m ;
bin4..          p('4')       =l= y('4')*m ;
bin5..          p('5')       =l= y('5')*m ;
bin6..          p('6')       =l= y('6')*m ;
costpart1..     uc('1')      =g= (100-0.01*p('1'))*p('1') ;
costpart2..     uc('2')      =g= (100-0.01*p('2'))*p('2') ;
costpart3..     uc('3')      =g= (100-0.01*p('3'))*p('3') ;
costpart4..     uc('4')      =g= (100-0.01*p('4'))*p('4') ;
costpart5..     uc('5')      =g= (100-0.01*p('5'))*p('5') ;
costpart6..     uc('6')      =g= (100-0.01*p('6'))*p('6') ;
cost..          z            =e= sum(i,A(i)*y(i))+sum(i,uc(i))+sum(i,Inv(i)*h(i)) ;

model case /all/;

p.fx('5') =1400;

option optcr=0;
solve case using miqcp minimizing z;
display p.m, p.l, z.l, uc.l;
```

## APÊNDICE B - EXEMPLO DE RESULTADO GERADO PELO GAMS

---- VAR p  production in each period

    LOWER    LEVEL    UPPER    MARGINAL

1    .    1500.000    +INF    .
2    .    1500.000    +INF    .
3    .    600.000    +INF    .
4    .    .    +INF    .
5    .    400.000    +INF    .
6    .    1000.000    +INF    .

---- VAR Inv  final inventory in each period

    LOWER    LEVEL    UPPER    MARGINAL

1    .    .    +INF    3.000
2    .    .    +INF    3.000
3    .    200.000    +INF    .
4    .    .    +INF    6.000
5    .    .    +INF    3.000
6    .    .    +INF    3.000

        LOWER    LEVEL    UPPER    MARGINAL

---- VAR k    -INF    600.000    +INF    .

k  total holding cost

---- VAR y  binary in each period

         LOWER     LEVEL     UPPER    MARGINAL

1        .         1.000     1.000    2000.000
2        .         1.000     1.000    2000.000
3        .         1.000     1.000    2000.000
4        .         .         1.000   -2.980E+5
5        .         1.000     1.000    2000.000
6        .         1.000     1.000    2000.000

                   LOWER     LEVEL     UPPER    MARGINAL

---- VAR z         -INF     10600.000   +INF       .

  z  total cost

**** REPORT SUMMARY :      0   NONOPT
                           0   INFEASIBLE
                           0   UNBOUNDED

GAMS Rev 145  x86/MS Windows                11/20/13 15:22:57 Page 6
G e n e r a l   A l g e b r a i c   M o d e l i n g   S y s t e m
E x e c u t i o n

----      99 VARIABLE p.L  production in each period

**1 1500.000,   2 1500.000,   3  600.000,   5  400.000,   6 1000.000**

----    99 VARIABLE p.M  production in each period

        (  ALL      0.000 )

----    99 VARIABLE z.L             =    10600.000  total cost

EXECUTION TIME      =      0.000 SECONDS      3 Mb  WIN222-145 Apr 21, 2006

USER: GAMS Development Corporation, Washington, DC   G871201/0000CA-ANY
      Free Demo,  202-342-0180,  sales@gams.com,  www.gams.com   DC0000

46